\documentclass[manuscript]{aastex}
\usepackage{natbib}
\usepackage{graphicx}% Include figure files
\usepackage{dcolumn}% Align table columns on decimal point
\usepackage{bm}% bold math
\usepackage{multirow}
\usepackage{amsmath}
\usepackage{amsfonts}

\shorttitle{}
\shortauthors{}

\begin{document}

\title{Protoplanetary disk shadowing by gas infalling onto the young star AK~Sco. }
\author{Ana I G\'omez de Castro$^1$, Robert O.P. Loyd$^2$, Kevin France$^2$,  Alexey Sytov$^3$, Dmitry Bisikalo$^3$}
\author{Ana In\'es G\'omez de Castro}
\affil{AEGORA Research Group, Fac. de CC Matem\'{a}ticas, Universidad Complutense, 28040 Madrid, Spain}
\author{Robert O.P. Loyd }
\affil{Department of Astrophysical and Planetary Sciences, Laboratory for Atmospheric and Space Physics, University of Colorado, Boulder, CO 80309, USA.}
\author{Kevin France}
\affil{Department of Astrophysical and Planetary Sciences, Laboratory for Atmospheric and Space Physics, University of Colorado, Boulder, CO 80309, USA.}

\author{Alexey Yu. Sytov}
\affil{Institute of Astronomy of the Russian Academy of Sciences, Moscow, Russia}
\author{Dmitry Bisikalo}
\affil{Institute of Astronomy of the Russian Academy of Sciences, Moscow, Russia}
\begin{abstract}
Young solar-type stars grow through the accretion of material from the circumstellar disk during pre-main sequence (PMS) evolution. The ultraviolet radiation generated in this process plays a key role in the chemistry and evolution of young planetary disks. In particular, the hydrogen Lyman-$\alpha$ line (Ly$\alpha$) etches the disk surface by driving photoevaporative flows that control disk evolution. Using the Hubble Space Telescope, we have monitored the PMS binary star AK~Sco during the periastron passage and have detected a drop of the H$_2$ flux by up to 10\%  lasting 5.9 hours. We show that the decrease of the H$_2$ flux can be produced by the occultation of the stellar Ly$\alpha$ photons by a gas stream in free fall
from 3~R$_*$. 
Given the high optical depth of the Ly$\alpha$ line, a very low gas column of $N_H > 5 \times 10^{17}$~cm$^{-2}$ 
suffices  to block the Ly$\alpha$ radiation without producing noticeable effects in the rest of the stellar spectral tracers.

\end{abstract}
%--------------------------------------------------------------------------------------------------------------------

\section{Introduction}

Solar-like pre-main sequence stars or T Tauri stars (TTSs) are complex dynamical systems made of two basic components, star and accretion disk. Stellar magnetospheres play a key role as dissipative interfaces between the star and the disk (G\'omez de Castro 2013). They absorb and reprocess part of the angular momentum excess of the infalling material and channel the flow into the open holes of the stellar magnetic configuration (Ghosh \& Lamb, 1979, Konigl 1991,  Romanova et al. 2012). Ultraviolet (UV) radiation from the magnetosphere-atmosphere-outflow ensemble is strong in TTSs, reaching values up to 50 times the main sequence luminosities (G\'omez de Castro \& Marcos-Arenal 2012, Ardila et al. 2013). This UV excess has a profound impact on the evolution of young planetary disks.

The lifetime, spatial distribution, and composition of gas and dust of young (age $< 30$ Myr) circumstellar (CS) disks are important properties for understanding the formation and evolution of extrasolar planetary systems. Disk gas regulates planetary migration (Armitage et al. 2002, Trilling et al. 2002) and the migration timescale is sensitive to the specifics of the disk surface density distribution and dissipation timescale (Armitage et al. 2007). Moreover, the time available for planetary embryos to coalesce cores and accrete gaseous envelopes is strictly limited by the 1-10 Myr lifetime of their parent disk (Alexander \& Armitage, 2009, Hernandez et al., 2007).

Some 80\% of the FUV photons incident on the disk surface are  Ly$\alpha$ (Herczeg et al. 2004, Schindhelm et al. 2012) and CS disks are comprised primarily of H$_2$ molecules that efficiently absorb the Ly$\alpha$ photons, producing
the subsequent electron cascade 
and radiation that serves as the most sensitive diagnostic of the molecular gas in the disk surface (e.g., Herczeg et al. 2006, France et al. 2012). H$_2$ emission is observed in all accreting TTSs, including those with transitional disks. In some sources H$_2$ emission is produced in the outflow but in most of them emission originates within the innermost 3 AU of the disk (France et al. 2012). The H$_2$ flux decreases as disks lose their gas. Gas removal is driven by the absorption of stellar UV photons that heat the gas leading to the generation of photoevaporative flows; however, models are uncertain at the order of magnitude level (Gorti et al. 2009, Alexander et al. 2014).  The comparison between theoretical predictions and actual observations reinforce this evaluation. The dust disk clearing timescale is expected to be 2-4 Myr (Alexander et al. 2006, Hernandez et al., 2007), however recent results indicate that inner molecular disks can persist to ages $\sim 10$~Myr in TTSs (France et al. 2012, Ingleby et al. 2011, Salyk et al. 2009).

A major source of uncertainty is the impact of obscuration by the accretion flow on disk irradiation. Ly$\alpha$ photons are primarily produced by the stellar/magnetosphere complex and therefore, they can be absorbed by the infalling neutral gas.  To date, no observations have tracked the propagation of the accretion flow and the subsequent level of disk irradiation.  In a generic TTS, it is unfeasible to separate a single accretion event; accretion occurs at a rather regular pace. However, this is not so in some types of PMS close binaries; systems composed of two equal mass stars in highly eccentricity orbits experience a significant enhancement of mass infall during the periastron passage (G\'omez de Castro et al. 2013, hereafter Paper~I). Therefore, accretion events are predictable and can be properly timed. 
Recognising that such systems could be used as a laboratory to study the effect of accretion flows on disk irradiation, our team undertook observations of the AK Sco system. AK~Sco is composed of two F5 type stars in an eccentric orbit that get as close as 11 stellar radii at periastron passage (see Table~1). The stellar atmospheres show evidence of the disturbances induced by the tide (G\'omez de Castro, 2009). 
The circumbinary disk is resolved in the infrared (H-band); the visibility profiles have been interpreted using a ring model and provide a radius of $0.58\pm 0.04$~AU for the ring that seems coplanar to the binary orbit (Anthonioz et al. 2015). These observations are in excellent agreement with the theoretical  model of the system (Paper~I); the high eccentricity of the orbit, together with the similar masses of both components creates a bar-like potential that distorts the inner part of the circumbinary disk  and creates an inner cavity of radius roughly three times the semimajor axis ($A$). 
Our team tracked the system during periastron passage in the UV with the Cosmic Origins Spectrograph (COS) instrument on board the Hubble Space Telescope.  The spectral range was selected to be sensitive to the presence of molecular gas close to the star (H$_2$), as well as to atomic plasma radiation, produced by the accretion flow, that excites the molecular lines. 

\section{Observations and analysis}

AK~Sco was tracked during the periastron passage (phase 0.992 to 1.023) in August 2014, using gratings G130M and G160M (R=17,000). 
Bright H$_2$ fluorescent emission lines were detected, similar to all other accreting classical TTSs (France et al. 2012). The lines have Gaussian-like symmetric profiles with suprathermal widths (Gaussian FWHM ranges from 63-69 km s$^{-1}$) implying that the line broadening is dominated by macroscopic motions. The line centres remained approximately constant over the course of the observations. The small observed variations (10 km s$^{-1}$) can be attributed to zero-point uncertainties in the COS wavelength solution introduced by the re-acquisition of AK~Sco on subsequent orbits.
To study the temporal evolution of the H$_2$ emission near periastron, we focused on four lines from the brightest H$_2$ progression, the (1 - 6) R(3) 1431 \AA\, (1 - 6) P(5) 1446 \AA\, (1 - 7) R(3) 1489 \AA\, and (1- 7) P(5) 1504 \AA\ emission lines of the [1,4] progression. The lines are consistent with an origin in an optically thin hot (T$ \sim 2500$~K) molecular gas without significant reddening effects across the 75 \AA\ spanned by these lines. 

The H$_2$ light curve is calculated in two ways to control for systematic effects introduced by the choice of analysis technique. The spectra were subdivided into time intervals of approximately 360 seconds and the four lines of interest isolated. First, the continuum-subtracted line fluxes were simply integrated from $-80 - +80$ km s$^{-1}$. Second, as a complementary approach, we also performed Gaussian fitting of each of the time-resolved one-dimensional line profiles individually, following the procedure outlined in France et al.( 2012) to account for the local continuum and the shape of the HST line-spread-function that is fed to COS. All four lines were fitted at each time step. The individual H$_2$ line fluxes were divided by the theoretical branching ratios to give the total flux in the [1,4] progression; the average velocity and total progression flux were measured with the uncertainty defined as the standard error on the mean of these measurements.  The low S/N of the individual H$_2$ line profiles in the 360 seconds cadence data prevent robust measurements of the time-resolved line widths.
We estimated uncertainties beginning with the assumption of Poisson statistics for the total counts within signal and background bins and propagating from there. We propagated these errors through the background subtraction that yielded count rate measurements. For the continuum subtraction, we used a maximum-likelihood method assuming Gaussian errors to compute the fit and estimate associated errors on fit parameters. These fit uncertainties are then propagated through the continuum subtraction. For absolute flux measurements, we neglect any flux calibration error. The absolute flux calibrations of COS and STIS are expected to be accurate within 5\% (Bostroem. and Proffitt, 2011; Debes et al., 2015).
We show the light curves from both flux measurement approaches as the orange and grey points, respectively, in Figure~1 (top, left panel).  We observe a significant decrease in the H$_2$ light curve during periastron passage, extending from approximately -0.003 - +0.015 (5.9 hours), with a maximum flux decrement of ~10\% at phase 0.002. 

Continuum subtracted fluxes have been computed for some high SNR lines in the spectrum (C~IV, Si~IV, N~V, Si~III, C~III).
The drop detected in the H$_2$ flux is not observed in the rest of the tracers, neither in the spectral lines nor the continuum. Therefore the decrease in the H$_2$ light curve is unlikely to be caused by extinction effects such as the passage of a dusty cloud through the line of sight.

\section{Interpretation. Evidence of occultation by an infalling gas stream}

There are three possible causes of the drop in the H$_2$ flux:

\begin{itemize}

\item A drop in the Ly$\alpha$ flux.

\item A variation in the H$_2$  distribution.

\item Disk shadowing by transient gas close to the star that absorbs Ly$\alpha$ photons.
\end{itemize}

The H$_2$ time behaviour is markedly different from the behaviour of the atomic emission lines observed simultaneously by COS, which are flat at phase $\leq 0.000$ and then rise by $10–15$\% between 0.002 and 0.024 (see Figure 1). Only C~IV shows a markedly different behaviour with a significant rising at the time the H$_2$ flux decreases. Therefore, the H$_2$ flux decrement cannot be associated with a drop of the  intrinsic  stellar Ly$\alpha$ flux. 

Another possibility is that the total surface of H$_2$ molecules collecting Ly$\alpha$ photons has decreased due to the enhancement of accretion at periastron. As the Ly$\alpha$ flux pumping the H$_2$ molecules varies as $r^{-2}$, one might expect that a fast re-distribution of CS  matter could account for the 10\% variability.  However, the numerical simulations run by our team show a rather stable configuration around the periastron passage. In the simulations, the gas flow is described by the Euler equations assuming an adiabatic equation of state 
(see Paper~I for full details); a constant accretion rate of $0.5 \times 10^{-9}$~M$_{\odot}$~yr$^{-1}$  through the circumbinary disk is assumed in agreement with the thresholds derived from the infrared observations of the disk (Alencar et al. 2003). 
The bar-like potential produced by the orbit distorts the inner part of the circumbinary disk and evacuates an inner cavity of radius roughly three times the semimajor axis ($A$). The system works like a gravitational piston, matter is dragged into the cavity mainly during the apastron leading to the formation of circumstellar disks around each component. At periastron, the circumstellar structures contact each other leading to an accretion outburst; the accretion rate onto the stars varies with the orbit. During the short time lapse monitored with HST, numerical simulations predict a variation of the accretion rate by a factor of $\sim 5$ (see Fig.~2) however, the mass flow proceeds along the CS spiral structures without significantly altering the overall molecular gas distribution. This is shown in Fig.~3, where the variation of the geometric cross section of the CS matter to Ly$\alpha$ photons  within the disk cavity (inner $3A$)  is plotted. This cross section $\Omega$ is calculated as,  
$$
\Omega = \frac {S_1(r_1) \cos (\theta )} {r_1^2} + \eta \frac {S_2(r_2) \cos (\theta )} {r_2^2} 
$$
where $S_1(r_1)$ being the surface area within radius $r_1$ and radius $r_1+\delta r$ around star~1 that contains gas to the numerical sensitivity of the code ($\rho =10^{-14}$~g~cm$^{-3}$) and $\theta$ is the incidence angle of the stellar radiation at $r_1$ ($\tan (\theta ) = 1.3 R_* /r_1$). The same definitions apply to star~2. The cross section  from both stars is added together, allowing for 
a balance factor, $\eta$, since Ly$\alpha$ luminosities and accretion rates may not be the same for both sources.  

Therefore, the source of the variations is, most likely, the head of the accretion flow located very close to the star that covers a significant solid angle of the stellar surface (and the corresponding Ly$\alpha$ flux), partially shading the disk from Ly$\alpha$ photons. Notice that the filament density must be low (or it must be devoid of dust) since the drop in H$_2$ flux is not accompanied by a decrease in the flux of nearby spectral lines (see Fig.~1). Dust associated extinction would have affected all spectral tracers. 

Given the high optical depth of the Ly$\alpha$ line, thin clouds could be opaque to the Ly$\alpha$ photons while producing a negligible effect in the rest of the spectral tracers; hydrogen column densities as low as $5\times 10^{17}$~cm$^{-2}$ suffice to block the stellar Ly$\alpha$ flux. The 10\% drop in flux lasts 5.9~h which is compatible with the free-fall time of matter from 3~R$_*$ for the parameters of the numerical simulations. Note that on these scales the gas dynamics is dominated by the stellar gravitational field, as is clearly shown by the distribution of the CS matter in the simulations
(see Fig.~2).

The computational grid in our numerical code has a resolution of 0.2~$R_*$ and an inner boundary of radius 2R$_*$ around each star. As a result,
it does not reach the small scales inferred above. Therefore, we developed a simple model to track the flow of a mass stream within
these close distances  and produce simulated light curves as output. 
In doing so, we neglected magnetic effects. This is a reasonable assumption since  AK~Sco's components are F5 type and no signatures of strong magnetic fields 
have been detected (see Paper~I). We have assumed that the infalling gas is shaped as a cylindrical filament or stream  opaque to Ly$\alpha$ photons at all points. The column density of $N_H > 5\times 10^{17}$ cm$^{-2}$ required for near total opacity is easily satisfied by any gas filament in the CS environment according to our simulations. 

Our model uses the analytical solution of the two-body problem and adopts for the conserved constants, angular momentum per unit mass, $h$, and total energy per unit mass, $\epsilon$,  the values provided by the numerical simulations at 3~R$_*$. The kinematics is described by two components: rotation around the star ($V_{\theta}$) and radial infall ($V_r$) with,

$$
V_{\theta} = \frac {h}{r}
$$

\noindent
with $r$ representing the distance to the stellar center, and

$$
V_r = \sqrt{ 2 \large(\epsilon - \frac {V_{\theta}^2}{2} + \frac {GM_*}{r} \large)}  \hspace{2cm} [1]
$$
with $M_*$ representing the stellar mass and G the gravitational constant; the mass of the filament has been considered negligible compared
to the stellar mass.
Angular momentum conservation during the fall drives the azimutal motion and makes the infalling stream to occult an increasing fraction of the stellar surface while falling. This simple model fits the data well, as displayed in Fig.~4; in the top panel the filament trajectory is drawn and in the bottom panel the best fitting light curve is superimposed to the observed light curve. The model light curve corresponds to the occultation of the stellar disk by a filament of length $\sim$3~R$_*$.  The  time is 15,330~s and from end to end, the trajectory subtends 1.37 radians on the stellar surface assuming that matter falls onto the stellar equator. Notice that the light curve is slightly asymmetric since the curvature of the  trajectory increases as matter approaches the stellar surface. This slight asymmetry is also noticeable in the data. 

In the model, two extreme assumptions have been tested concerning the solid angle subtended by the filament; they are represented with solid and dashed lines in Fig.~4. The solid line takes into account that the shadow produced by the filament depends  on the distance to the star since the width of the shadow projected by the filament on the surface increases as the filament approaches the surface.  The dashed line model represents also the shadowing but assuming that the filament transverse section decreases as it approaches the star as $r^{-2}$. This is a reasonable assumption if the gas were trapped in a flux tube within which the continuity equation is satisfied. The curves are significantly different, pointing out that the light curve can be used to characterise the properties of the flow.  Our observations are best fitted 
by the first option as shown in Fig.~4. Note that mass infall is assumed to proceed nearly simultaneously in both sources as in Fig.~2. 

From the maximum depth of the H$_2$ absorption in the light curve, the angular size of the filament, perpendicular to the trajectory, can be determined. Shading the Ly$\alpha$ emission by a $\sim 10$\% requires that a total solid angle of 0.628~rad$^2$ is occulted by the diffuse infalling gas. Since the extent of the filament along the trajectory is known (and well determined by the duration of the occultation), the extent perpendicular to the trajectory is estimated to be 0.32~rad (18.$^o$1). If the filament is assumed to be cylindrical, the shock on the stellar surface would cover 1.2\% of the visible hemisphere at the impact point; since the star rotates at a rate of $vsini = 18.5 \pm 1.9$~km~s$^{-1}$ (or $1.81\times 10^{-5}$~s$^{-1}$) the total shocked surface is slightly larger because the head of the filament falls 0.14~rad ahead of the 
tail on the stellar surface.  This size of the accretion spot is in good agreement with measurements for other PMS stars (Donati et al., 2008, Ingleby et al. 2013), though probably represents an upper limit since close to the surface the action of the magnetic field will channel the flow into thin shells or curtains of falling particles. 

Finally, note that the UV radiance at the point of impact must increase by a factor of $\sim 10$ to account for the 10-20\% (depending on the tracer) increase  of the UV emission (line and continuum) following the end of the H$_2$ flux drop (see light curves in Figure~1).  The Ly$\alpha$ line is not seen to rise with the rest of the tracers but this is an expected behaviour given the high optical depth of the line (see {\it i. e.} France et al. 2014) and the theoretical predictions for UV radiation from accretion shocks (Calvet \& Gullbring, 1998, Gomez de Castro \& Lamzin, 1999).

\section{Conclusions}

We have observed the highly eccentric, pre-main-sequence binary system AK Sco during the very close (11 stellar radii) passage of its two F5 component stars at periapsis. Coincident with this passage, we measured a marked 10\% decrease in the H$_2$ emission from  the disk and determined that the best explanation for this flux drop is the passage of an accreting filament of gas that shades some of the H$_2$ molecules on the disk surface from the stellar Ly$\alpha$ photons. 

This phenomenon ought to be common to all accreting PMS stars. Matter falling on the stellar surface is expected to occult a significant fraction of the stellar Ly$\alpha$ photons. As $\sim 80$\% of the FUV flux is radiated in the Ly$\alpha$ line in  solar-like PMS stars, this occultation has profound implications for the heating of the disk atmosphere and, hence, its evolution. Given the clear occultation of Ly$\alpha$  photons generated near the star in the moderate accretion rate AK~Sco system, one imagines that infalling matter in more rapidly accreting systems might completely attenuate the stellar Ly$\alpha$ flux irradiating the disk. In such cases, additional sources of FUV illumination, such as jets and outflows, may play an important role in the photoevaporative disk evolution.

\acknowledgments
The team thanks the Hubble Space Telescope support team for their assistance in the scheduling of the monitoring program that made these challenging observations feasible. 
AIGdC acknowledges financial support by the Spanish Government under grant AYA2011-29754-C03-01.  
K.F. and R.O.P.L. acknowledge support from the HST Guest Observer program (HST-GO-13372).
A.S. and D.B. acknowledge support by the Russian Science Foundation (Project nr. 15.12.30038).

{\it Facilities:} \facility{HST (COS)}.

%%%%FIGURES

\newpage

\begin{figure}[h]
\centering
\begin{tabular}{cc}
\includegraphics[width=7cm]{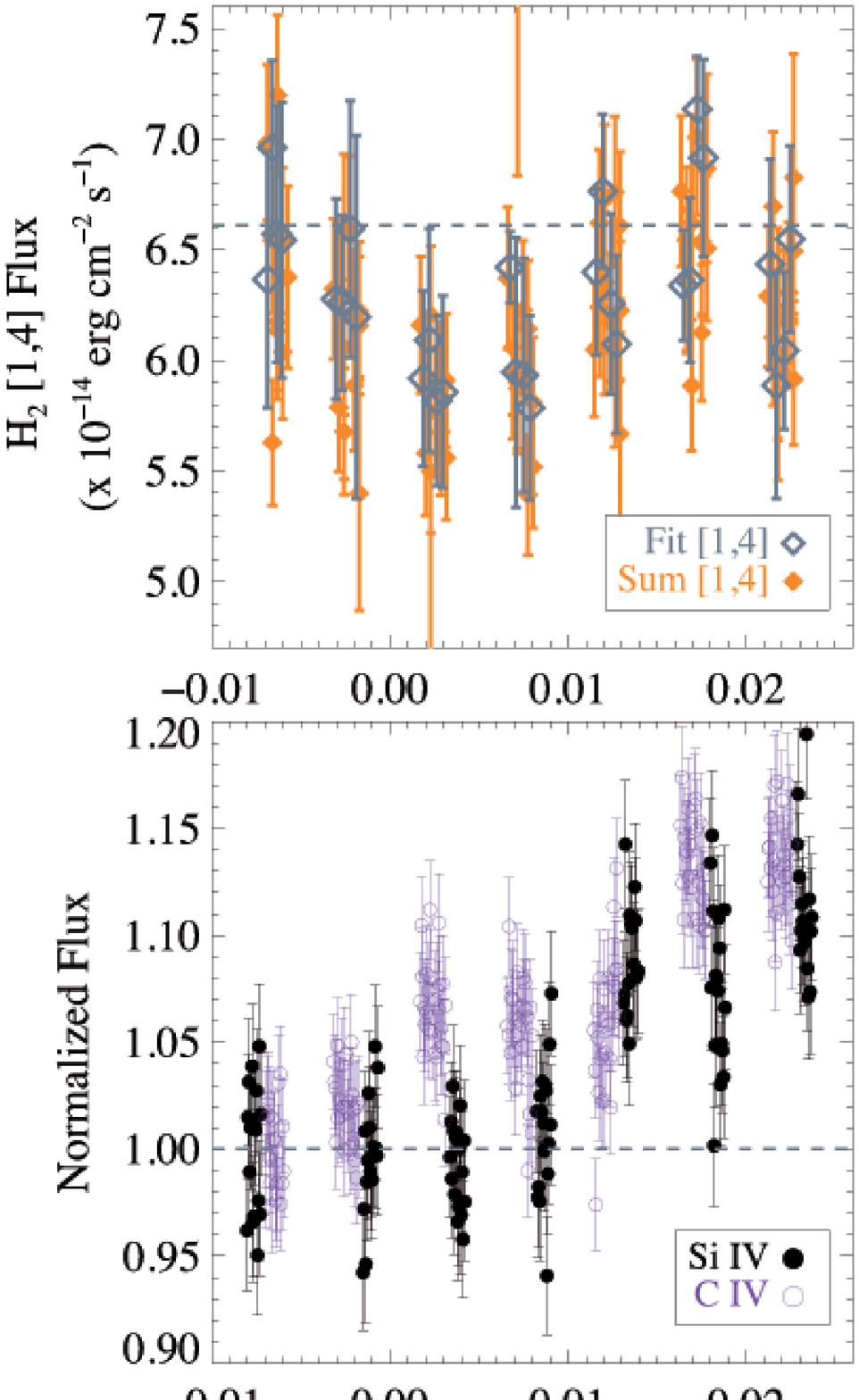} & \includegraphics[width=7cm]{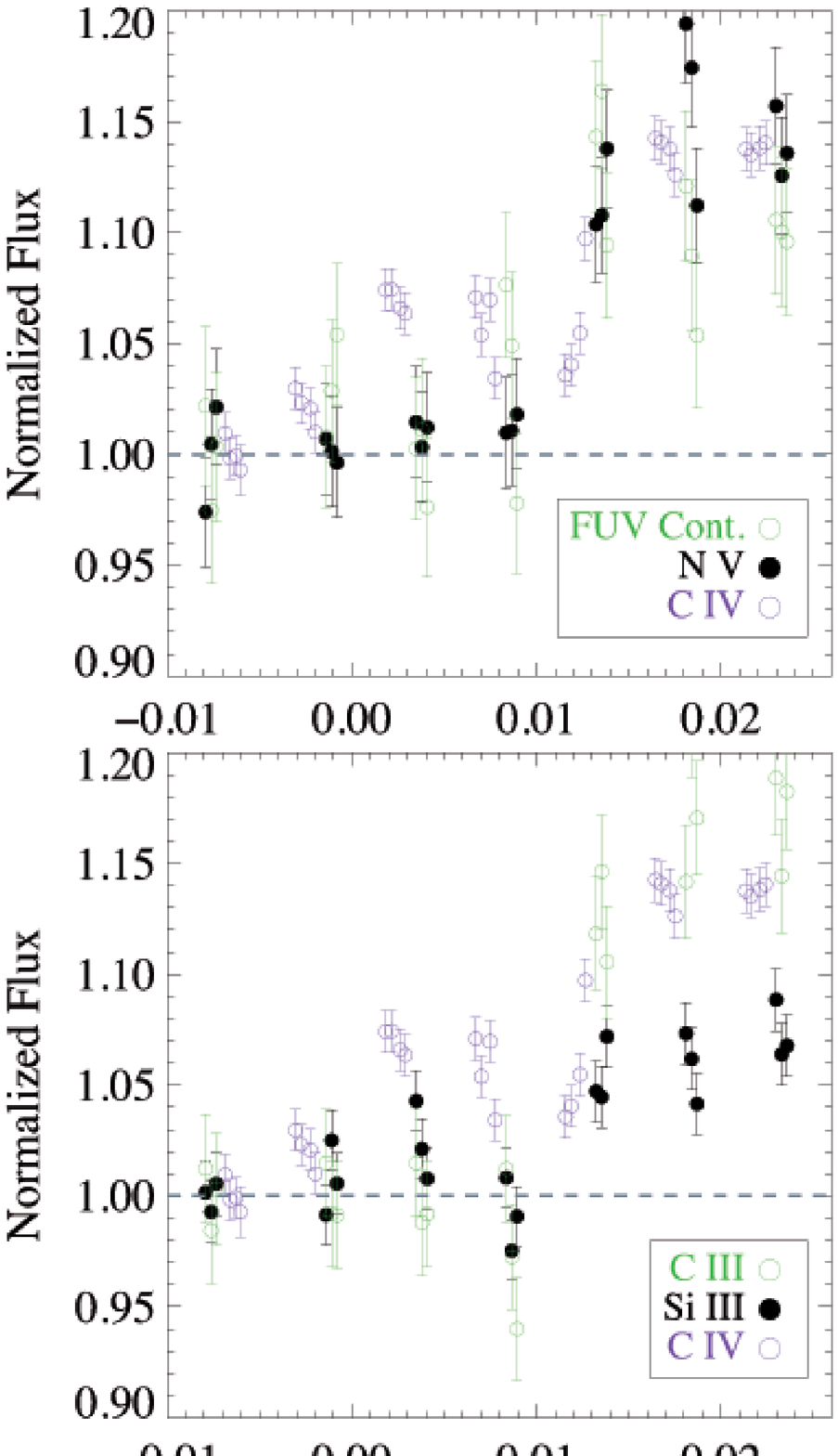} \\
\end{tabular}
\caption{Upper upper panel: H$_2$ flux variation during the periastron passage. Rest of panels: light curves of various UV lines during the periastron passage. UV lines are tracers of the mass accretion rate (Ardila et al. 2013, G\'omez de Castro \& Marcos-Arenal, 2012). }
\label{fig:H$_2$}
\end{figure}

\newpage

\begin{figure}[h]
\centering
\begin{tabular}{cc}
\includegraphics[width=7cm]{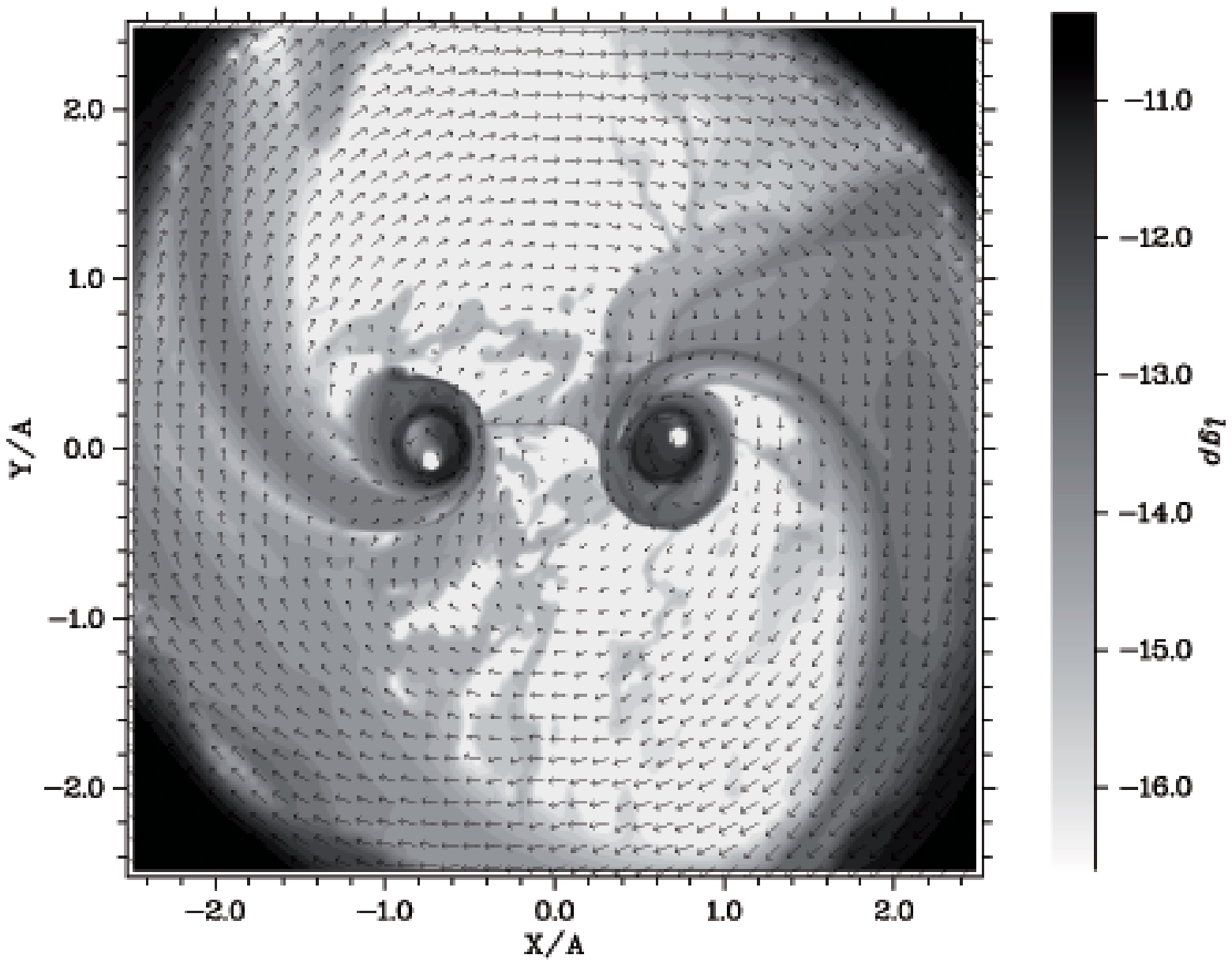}& \includegraphics[width=7cm]{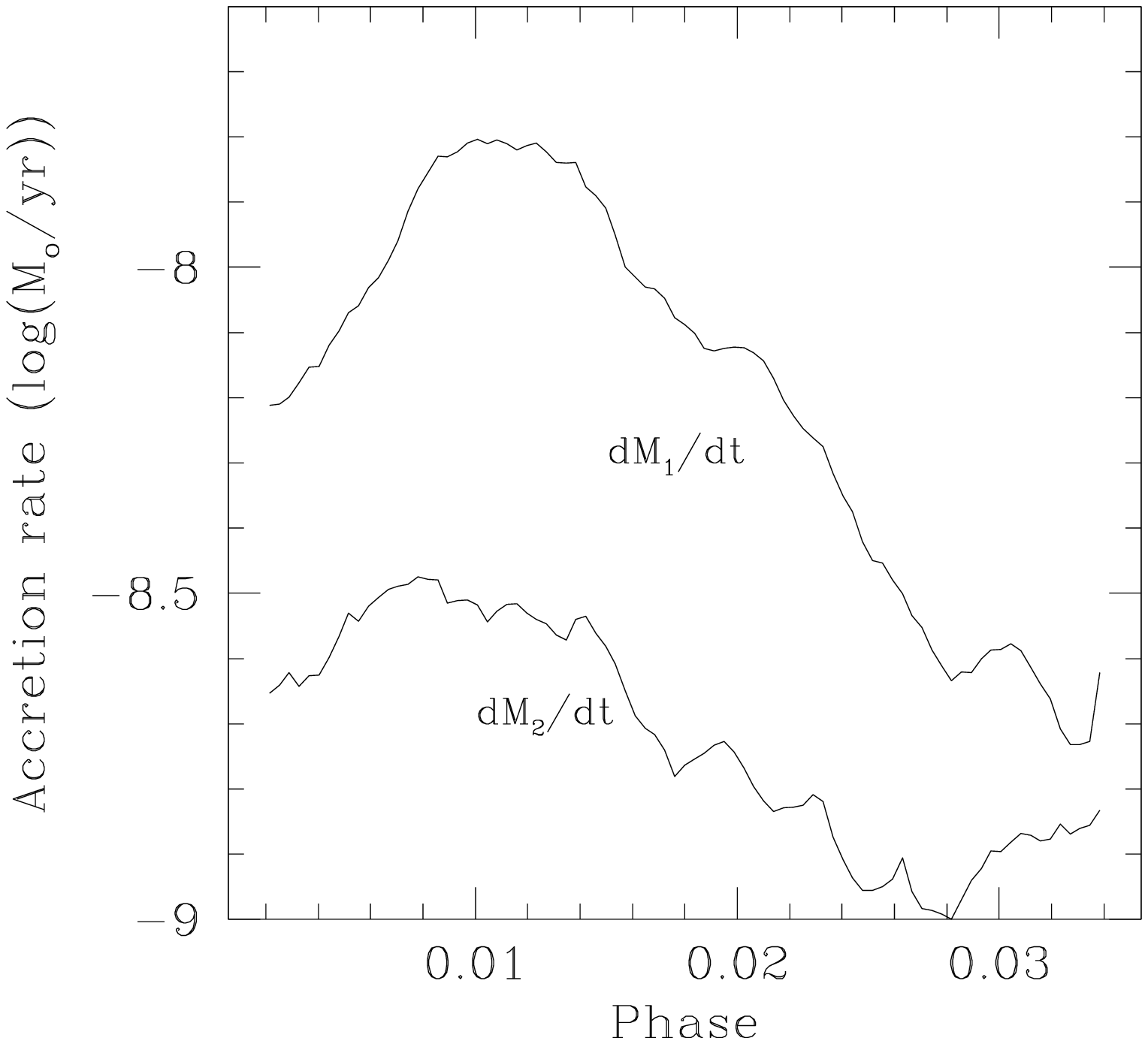} \\
\end{tabular}
\caption{ {\bf Left:} Expected distribution of the circumstellar matter distribution during periastron passage for a typical orbit from the numerical simulations of the system (from Paper I).
{\bf Right:} Predicted variation of the accretion rate onto the primary and the secondary during the Hubble monitoring for a sample orbit. }
\label{fig:mosaic}
\end{figure}

\begin{figure}[h]
\centering
\includegraphics[width=15cm]{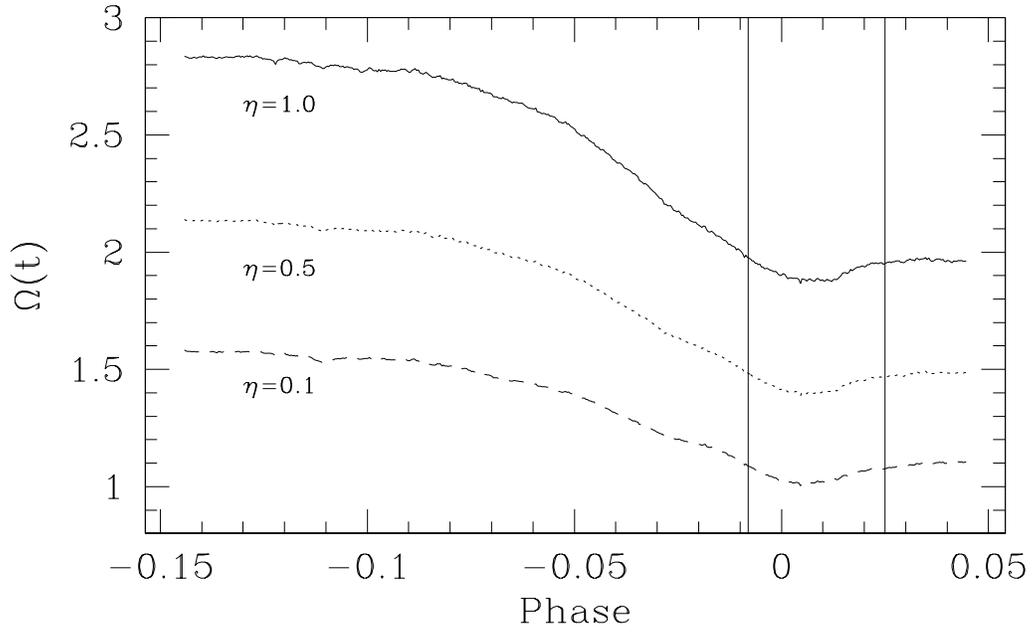}
\caption{Variation of geometrical cross section of the CS matter, $\Omega$, to Ly$\alpha$ photons, for $\eta = 1, 0.5$ and $0.1$. The vertical solid lines mark the limits of the phase interval monitored with HST.  }
\label{fig:mosaic}
\end{figure}

\newpage

\begin{figure}[h]
\centering
\includegraphics[width=15cm]{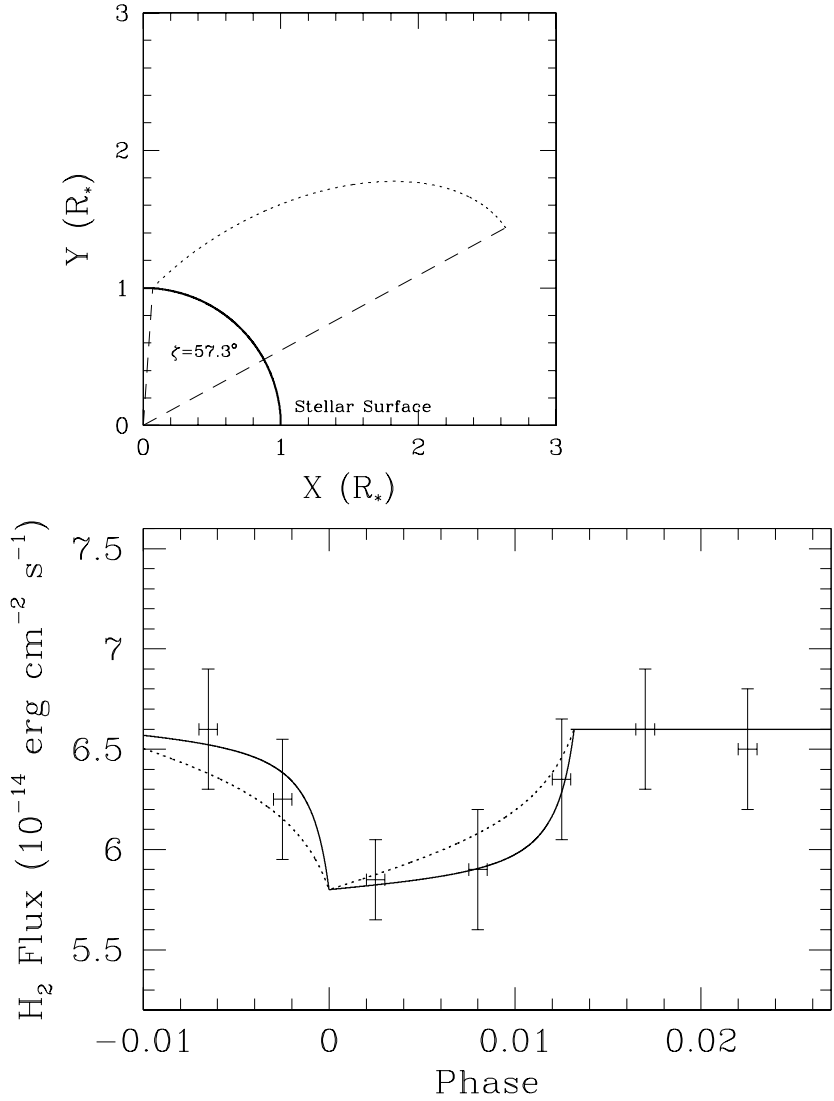}
\caption{{\it Top:} Trajectory of the infalling gas filament onto the star as predicted by equation [1]; the trajectory (dotted)
subtends 57.$^o3$ on the stellar surface (limits marked with dashed lines). {\it Bottom:} Light curve produced by the screening of stellar Ly$\alpha$ photons by the infalling gas. The best fitting model assumes that the filament section is kept constant during the fall (solid line); 
the asymmetry  is well reproduced by the infall trajectory. If the filament section decreases while apparoaching the stellar surface as $r^{-2}$, the light curve becomes more symmetric and shallower (dashed line). 
H$_2$ flux measurements are overlaid. }
\label{fig:model}
\end{figure}

\newpage
\begin{table*}
\caption{AK Sco main parameters.} %$^{(a)}$.}
\begin{center}
\small
\begin{tabular}{lll}
\hline
Property & Value & Source \\
\hline
Projected semimajor axis & $a\sin i = 30.77 \pm 0.12$R$_{\odot}$ & Andersen et al 1989 \\
Eccentricity & e= 0.47 & Andersen et al 1989, Alencar et al. 2003 \\
Orbital period & P=13.609453$\pm$ 0.000026 d & Andersen et al 1989, Alencar et al. 2003 \\
Periastron Passage & T=2,446,654.3634 ± 0.0086 &  Alencar et al. 2003 \\
Inclination & $ i=65^o-70^o$ & Alencar et al. 2003 \\
Age & 10-30 Myrs & Alencar et al. 2003  \\
Spectral type & F5 & Alencar et al. 2003\\
Stellar Mass & $M_* = 1.35 \pm 0.07$M$_{\odot}$ & Alencar et al. 2003\\
Radius & $R_* = (1.59 \pm 0.35) R_{\odot}$ & Alencar et al. 2003 \\
Projected rotation velocity & $v \sin i = 18.5 \pm 1.0$ km s$^{-1}$ & Alencar et al. 2003 \\ \hline
\end{tabular}
%\begin{tabular}{ll}
%{\bf (a) & Recent re-evaluations of the total stellar masses (Czekala et al. 2015) and the orbital elements
%and components masses (Anthonioz et al. 2015) are available and agree with these values.}
%\end{tabular}
\end{center}
\label{tab:table1}
\end{table*}


\begin{thebibliography}{}

\bibitem[alencaretal2003]{Alencar et al. 2003}
Alencar, S.H.P., Melo, C. H. F., Dullemond, C. P., Andersen, J., Batalha, C.,  et al.,  2003. A\&A, 409, 1037-1053 .
\bibitem[alexander et al. 2006]{Alexander et al. 2006}
Alexander, R. D., Clarke, C. J., Pringle, J. E., 2006. M.N.R.A.S, 369, 229-239.
\bibitem[aa2009]{aa2009}
Alexander, R.D., Armitage, P.J., 2009. ApJ, 704, 989-1001. 
\bibitem[alexandreaetal2014]{aetal2014}
Alexander, R., Pascucci, I., Andrews, S., Armitage, P., Cieza, L., 2014, Protostars and Planets VI, Henrik Beuther, Ralf S. Klessen, Cornelis P. Dullemond, and Thomas Henning (eds.), University of Arizona Press, Tucson, 914, pp. 475-496 . 
\bibitem[aetal89]{aetal1989}
Andersen, J., Lindgren, H., Hazen, M.L., Mayor, M., 1989. A\&A, 219, 142-150.
\bibitem[anthoniozetal2015]{ant2015}
Anthonioz, F., Menard, F., Pinter, C. et al., 2015, A\&A, 574A, 41. 
\bibitem[ardilaetal2013]{ardilaetal2013}
Ardila, D.R., Herczeg, G. J., Gregory, S.G., Ingleby, L. , France, K. et al.,  2013. ApJ, 207, 1-43.
\bibitem[armitageetal2002]{armitageetal2002}
Armitage, P.J., Livio, M., Lubow, S.H., Pringle, J.E., 2002. M.N.R.A.S, 334, 248-256.
\bibitem[armitage20070]{Armitage2007}
Armitage, P.J., 2007. ApJ, 665,1381-1390.
\bibitem[calvetetal98]{calvetetal98}
Calvet, N. Gullbring, E., 1998, ApJ, 509, 802.
\bibitem[Debesetal2015]{debesetal2015}
Debes, J., et al. 2015, Cosmic Origins Spectrograph Instrument Handbook, Version 7.0 (Baltimore: STScI)
\bibitem[franetal2012]{francetal2012}
France, K., Schindhelm, E., Herczeg, G. J., Brown, A., Abgrall, H., et al. , 2012. ApJSS, 756, 171.
\bibitem[franceetal2014]{franceetal2014}
France, K., Schindhelm, E., Bergin, E., Roueff, E., Abgrall, H. 2014, ApJ, 784, 127.
\bibitem[donatietal2008]{donatietal2008}
Donati, J.-F., Jardine, M. M., Gregory, S. G., Petit, P., Paletou, F. et al., 2008. M.N.R.A.S,  386, 1234.
\bibitem[goshandlamb]{goshandlamb}
Ghosh, P., Lamb, F.K., 1979. ApJ, 232, 259-276.
\bibitem[gdc2009]{gdc2009}
G\'omez de Castro, A.I., 2009. ApJ, 698, L108.
\bibitem[gdc2013a]{gdc2013a}
G\'omez de Castro, A.I., 2013. Planets, Stars and Stellar Systems Vol. 4, Oswalt, Terry D.; Barstow, Martin A., (Springer Science+Business Media Dordrecht, 2013), pp. 279-335.  
\bibitem[gdclamzin]{gdclamzin}
G\'omez de Castro, A.I., Lamzin, S.A., 1999, MNRAS, 304, 41.
\bibitem[gdcma]{gdcma}
G\'omez de Castro, A.I., Marcos-Arenal, P., 2012. ApJ., 749-764.
\bibitem[gdcetal2013]{gdcetal2013}
G\'omez de Castro, A.I., López-Santiago, J.L., Talavera, A., Sytov, A., Bisikalo, D.,  2013. ApJ,  766, 62.
\bibitem[gortietal2009]{gortoetal2009}
Gorti, U., Dullemond, C. P., Hollenbach, D., 2009. ApJ, 705, 1237-1251.
\bibitem[herczegetal2004]{hecrzegetal2004}
Herczeg, G. J., Wood, B. E., Linsky, J. L., Valenti, J. A., Johns-Krull, C. M., 2004. ApJ, 607, 369.
\bibitem[herczegetal2006]{herczegetal2006}
Herczeg, G. J., Linsky, J. L., Walter, F. M., Gahm, G. F., Johns-Krull, C. M., 2006. ApJSS, 165, 256.
\bibitem[hernandezetal2007]{hdzetal2007}
Hernandez, J., Calvet, N., Briceño, C., Hartmann, L., Vivas, A. K., et al. , 2007. ApJ, 671, 1784-1799.
\bibitem[inglebyetal2011]{inglebyetal2011}
Ingleby, L., Calvet, N., Bergin, E., Herczeg, G., Brown, A., et al. ,  2011. ApJ, 743, 105-116.
\bibitem[inglebyetal2013]{ingelbyetal2013}
Ingleby, L., Calvet, N., Herczeg, G., Blaty, A., Walter, F. et al., 2013. ApJ, 767, 112. 
\bibitem[k91]{k91}
Koenigl, A. , 1991. ApJ,  370, L39-L43.
\bibitem[romanovetal2012]{romanovaetal2012}
Romanova, M.M., Ustyugova, G.V., Koldoba, A.V., Lovelace, R.V.E., 2012. M.N.R.A.S, 421, 63-77.
\bibitem[salyketal2009]{salyketal2009}
Salyk, C., Blake, G. A., Boogert, A. C. A.,Brown, J. M., 2009. ApJ, 669, 340-347.
\bibitem[schindlemetal2012]{schindleetal2012}
Schindhelm, E., France, K., Herczeg, G. J., Bergin, E., Yang, H., Brown, A., et al., 2012. ApJ, 756, L23.
\bibitem[trillingetal2002]{trillingetal2002}
Trilling, D. E., Lunine, J. I., Benz, W., 2002, A\&A, 394, 241-251.

\end{thebibliography}
\end{document}